# Direct measurement of coherent subterahertz acoustic phonons mean free path in GaAs

R. Legrand,[1] A. Huynh,[1] B. Jusserand,[1] A. Lemaître[2] and B. Perrin[1*]

[1]Sorbonne Universités, UPMC Univ Paris 06, CNRS-UMR 7588, Institut des NanoSciences de Paris, F-75005 Paris, France
[2]Laboratoire de Photonique et de Nanostructures, CNRS., Route de Nozay, 91460 Marcoussis, France

Abstract:

Phonon mean free path is generally inferred from the measurement of thermal conductivity and we are still lacking precise information on this quantity. Recent advances in the field of high frequency phonons transduction using semiconductor superlattices give the opportunity to fill this gap. We present experimental results on the attenuation of longitudinal acoustic phonons in GaAs in the frequency and temperature ranges 0.2-1THz and 10-80K respectively. Surprisingly, we observe a plateau, in the frequency dependence of the attenuation, above 0.4THz, that we ascribe to a breakdown of Herring processes.

A large renewed interest appeared these last years in theoretical and experimental studies of phonons mean free paths (MFP) in semiconductors. Thermal transport in optoelectronic and microelectronic devices and engineering of new thermoelectric systems are strongly dependent on this parameter which is still not very well-known. A series of recent experiments using thermal conductivity spectroscopy technique [1-5] showed that phonons contributing to thermal conductivity have a large distribution of MFP. It can be suspected that low frequency phonons may have MFP in the micrometric range at room temperature and thus could play a large role in heat transport in nanosystems [6-8]. On the other hand, first-principles calculations [9-11] proved that the simplistic kinetic theory based on an averaged phonon MFP value can be very misleading. In this context accurate and direct MFP measurements for well-defined individual subterahertz phonon modes would be useful.

In this letter, we present experimental measurements of the attenuation of coherent longitudinal acoustic waves propagating along [100] with frequencies going from 0.2 to 1 THz in the temperature range 10-80K. Direct phonon MFP measurements in the subterahertz range cannot be done with a good accuracy by inelastic neutron and X-ray scattering and is out of reach for standard light scattering methods. Picosecond ultrasonics methods combined with metallic films as transducers have been used up to 0.1 THz for MFP determinations in bulk samples [12-15] and a few hundreds of GHz in strongly absorbing amorphous thin films [16-17]. Recently semiconducting superlattices (SL) proved to be very efficient emitters and detectors of subterahertz coherent acoustic waves [18-25] and have also been used for phonons MFP measurements [26-28]. Excited by a femtosecond laser pulse with an energy above the electronic band gap, a SL can emit a discrete set of frequencies determined by its period superimposed to a low frequency spectrum extending up to a few tens of GHz. The discrete frequencies correspond to the lower edges of phonon energy gaps at the center of the Brillouin zone $q_{ac} = 0$, where $q_{ac}$ is the acoustic wave vector. These SLs may also resonantly detect any mode satisfying the selection rule $q_{ac} = k_i + k_d$, where $k_i$ and $k_d$ are the incident and scattered electromagnetic wavevectors respectively. Furthermore, the detection is strongly enhanced if the probe laser wavelength is tuned closed to a SL electronic transition. In spite of this wavevector mismatch between emission and detection, it is nevertheless possible to emit and detect phonons using a single SL configuration but this is not the optimal way to perform subterahertz acoustic experiments. We used four different configurations: i) reflection experiments on a single SL which acts both as a phonon emitter and detector, ii) reflection experiments on a single SL embedded in an optical cavity which strongly modifies the selection rule $q_{ac} = k_i + k_d$, [29-31], iii) reflection experiments on a large period SL such that $k_i + k_d = 0$, (mod $2\pi / d$), and lastly iv) transmission experiments with slightly different SLs deposited on the opposite sides of a wafer in order to satisfy simultaneously the emission and detection selection rules.

Our experiments have been performed for 9 different frequencies reported in table I. Below 0.4 THz the amplitude of the acoustic waves having made a round trip in the sample has been measured as a function of temperature. Above 0.4 THz, the reflection coefficient of acoustic waves on the sample back-side can be strongly affected by its quality and cleanness. To avoid such problems, the acoustic attenuation for the highest frequencies has been obtained recording the high frequency component transmitted through the sample and detected with a SL deposited on the wafer back-side [23].

| Frequency (THz) | Config | Propagation distance (μm) |
|---|---|---|
| 0.202 | RT-large per. | 750 |
| 0.243 | RT – LP[a] | 750 |
| 0.292 | RT – cavity[b] | 740 |
| 0.356 | RT | 692 |
| 0.394 | Trans/RT | 360/692/983 |
| 0.714 | Trans | 368 |
| 0.786 | Trans | 361 |
| 0.819 | Trans | 361 |
| 1.008 | Trans | 368 |

TABLE I. Different configurations used for the experiments. "RT" is for round trip, "Trans" for transmission.
[a]SL with a large period such that $\boldsymbol{k}_i + \boldsymbol{k}_d = \boldsymbol{0}, (\mathrm{mod}\ 2\pi/d)$.
[b]SL embedded in an optical cavity.

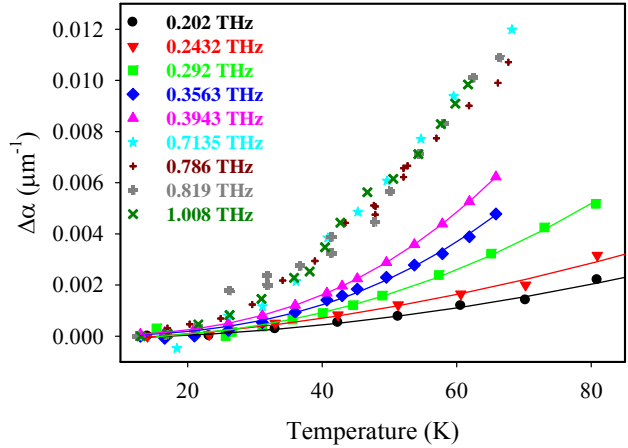

FIG. 1(color online) Deviational sound absorption curves $\Delta\alpha(T)$. The colored full lines for frequencies up to 0.4 THz are guide for the eyes.

In our experimental configuration, we can only measure the relative change with temperature of the sound absorption coefficient Δα, defined by

$$\Delta\alpha(T) = \ln\big(A(T_0)/A(T)\big)/d, \quad (1)$$

where d is the propagation distance, $A(T_0)$ the amplitude of the discrete Fourier component at frequency $\nu_{ac}$ of the time resolved detected signal, at the lowest temperature we could achieve (about 10K), and $A(T)$ the same amplitude at temperature T. Experimental results are reported on figure 1. For frequencies up to 0.4 THz, the attenuation increases with temperature and frequency as can be expected. Surprisingly, above 0.4 THz, a plateau appears in the frequency dependence. In order to get rid of possible artifacts, complementary experiments were undertaken. Indeed this feature could also be attributed to i) nonlinear interactions between the discrete high frequencies of the generated spectrum and its lower continuous part. However, different pump laser powers were tested and no difference was observed. ii) or to the proximity of the probe wavelength to a SL electronic transition were photoelastic coefficients are very high. We indeed can suspect that the SL properties slightly change with temperature. This would affect the measured acoustic wave amplitude signal. To check this point several tests were carried out. First, we renormalized the amplitudes of high frequency components of the spectrum by the amplitude of low frequency (amplitude of echoes envelopes observed with an interferometric detection or Brillouin component at 40GHz) for which $\Delta\alpha$ is negligible on the investigated temperature range. In another set of measurements, the signal was optimized for each temperature by tuning the wavelength. In both cases, no significant difference was observed on the values of Δα. Last, a more crucial test was performed by comparing experiments done at the same frequency for three different propagation distances.

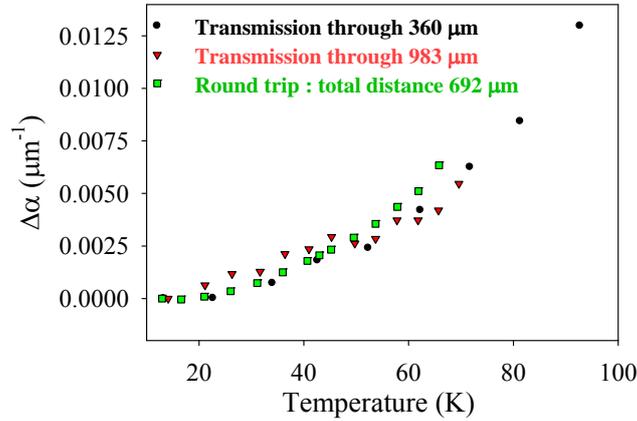

FIG. 2 (color online) Deviational sound absorption curves $\Delta\alpha(T)$ obtained for the same frequencies but on three samples and different propagation distances.

On figure 2 are compared results obtained at 0.4 THz for two transmission experiments through wafers with respective thickness of 360 and 983 μm and a round trip experiment done on a total distance of 692 μm. Any dependence of the emitters and detectors efficiencies on temperature would have had the most important effect for the shorter distances. As it can be seen on figure 2, the three sets of data obtained for the three different distances give the same results for Δα within an error bar of roughly 15%.

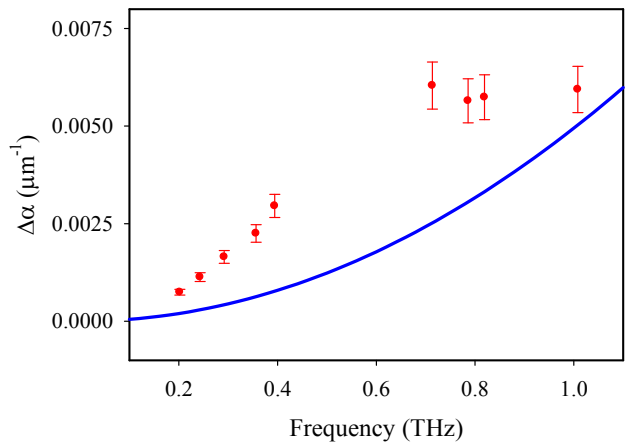

FIG. 3 (color online) Deviational sound absorption curves $\Delta\alpha(\nu_{ac})$ estimated for the same temperature 50K (red circles). The contribution of Herring processes estimated with the asymptotic formula of Simons (full blue line) departs strongly from the experimental data.

Having discarded most probable experimental sources of error, we have investigated the possible mechanisms responsible for this unexpected behavior with temperature. Two attenuation mechanisms have to be considered in the very low temperature limit. The spontaneous anharmonic decay of phonons gives a contribution $\alpha_{sd}(\mu m^{-1}) = 1.31 \times 10^{-4} \nu^5$, as estimated by Berke [32], where the acoustic frequency ν is expressed in THz. The second mechanism (when the dopant concentration is low) is the isotopic scattering which gives a contribution $\alpha_{iso}(\mu m^{-1}) = 7.5 \times 10^{-4} \nu^5$ [33]. Both contributions are rapidly overtaken by scattering processes and stimulated fission processes when temperature increases; thus the experimental values we obtained for Δα at the highest temperatures we achieved should be very close to the absolute value of the inverse phonon MFP and provide unique experimental data to be compared to existing models for acoustic attenuation.

Stimulated fission processes can contribute to the MFP temperature dependence above very low temperatures but this contribution remains small compared to scattering processes such as LA+TA→LA and LA+TA→TA. The latter process is forbidden when elastic isotropy is assumed but Herring [34] suggested that it should be very efficient in anisotropic systems. For such crystals, Herring showed that a longitudinal phonon, with energy $\omega_{ac}$ can interact with much higher-frequency transverse thermal phonons of different polarizations and wave vectors close to a symmetry axis along which there is degeneracy. A general expression for the sound absorption coefficient related to this process can be written as

$$\alpha_H(\omega_{ac}) = \frac{\hbar}{128\pi^2 \rho^3 v_l} \int \frac{|\Lambda_{q,q',q+q'}|^2 (n(\omega') - n(\omega' + \omega_{ac})) \delta(\omega_{ac} + \omega(q',st) - \omega(q+q',ft))}{\omega_{ac} \omega'(\omega_{ac} + \omega')} dq', \quad (2)$$

where $\Lambda_{q,q',q+q'}$ is a coupling parameter, $\rho$ the specific mass, $v_l$ the longitudinal sound velocity along [100], $q'$ the slow transverse thermal phonon wave vector and $\omega' = \omega(q',st)$ its energy ; the indices *st* and *ft* means slow and fast transverse respectively and $n(\omega')$ is the phonon population. Neglecting phonon dispersion, and using an elastic continuum approximation for the coupling parameter, Herring predicted an asymptotic behavior $B\nu^p T^{5-p}$ for $\alpha_H$ in the limit of low frequencies and low temperatures, where the exponent *p* depends on the crystal symmetry. Later on, Simons [35] derived an expression for the Herring processes contribution in cubic crystals: in that case $p=2$ and the prefactor B can be expressed in terms of second and third order elastic constants. The calculation for GaAs gives $\alpha_{Hasy.}(\mu m^{-1}) = 3.96 \times 10^{-8} \nu^2 T^3$. Using this expression, we show in figure 3 that this prediction strongly underestimates the attenuation at 50K up to 1 THz. This is not surprising since the asymptotic expression $\alpha_{Hasy.}$ holds for a very limited frequency and temperature range as we shall see now. The exact calculation of $\alpha_H$ could be achieved using a microscopic model or ab initio calculations for the phonon dispersion and coupling parameters. However calculation of Herring processes occurrence for subterahertz longitudinal phonons would require a very fine mesh of the Brillouin zone. This work is beyond the scope of this paper but we could learn interesting information about the frequency behavior of $\alpha_H$ by considering the behavior of the two phonon density of states for Herring processes defined by

$$\rho_H(\omega_{ac}) = \int_{BZ} \delta(\omega_{ac} + \omega(q',st) - \omega(q+q',ft)) dq'. \quad (3)$$

Let's define the spherical coordinates $(q',\theta,\varphi)$ of the wavevector $q'$. If we still neglect phonon dispersion, the energies $\omega(q',st)$ and $\omega(q+q',ft)$ can be rescaled by the acoustic energy $\omega_{ac}$. The calculation of $\rho_H(\omega_{ac})$ requires then the determination of the values $x = \omega(q',st)/\omega_{ac}$ which satisfy energy conservation $x(\theta,\varphi) = \omega(\mathbf{q}+\mathbf{q}',ft)/\omega_{ac} - 1$. The calculated density of these solutions $\rho_2(x)$ is given by

$$\rho_2(x) = \frac{x^2}{2\pi^2} \int_0^{2\pi} d\varphi \int_0^{\pi} \sin\theta \, \delta(x - x(\theta,\varphi)) \, d\theta. \tag{4}$$

This density is displayed on Figure 4 neglecting phonon dispersion and using $C_{11}=121.07$, $C_{12}=54.77$, $C_{44}=60.36$ GPa for GaAs elastic constants. We show that the asymptotic value of $\rho_2(x)$ restricted to $\mathbf{q}'$ wavectors close to the $A_4$ symmetry axes of a cubic system is given by

$$\lim_{x\to\infty} \rho_2(x) = \rho_{2,A_4}(x) = \frac{12}{\pi^2} \frac{C_{44}}{(C_{11}-C_{44})(A^2-1)} K\left(\frac{\sqrt{2A+1}}{A+1}\right) = 0.894, \tag{5}$$

where $A = (C_{44}+C_{12})/(C_{11}-C_{44})$ and $K$ is the complete elliptic integral of the first kind. We can see on figure 4 that the exact calculation departs rapidly from this asymptotic value as x decreases. For example, another important contribution appears at smaller x values due to transverse phonon close to $A_3$ axis; in that case, the constant-frequency surfaces in wave-vector space do not intersect like for $A_4$ axis but just touch. We show that the contribution of the $A_3$ axis to $\rho_2(x)$ is given by $\rho_{2,A_3}(x) = 7.62/x$ and is greater than $\rho_{2,A_4}(x)$ for $x<10$. For x = 20, the exact calculation gives

$$(\rho_2(x) - \rho_{2,A_4}(x))/\rho_{2,A_4}(x) = 80\% \tag{6}$$

The previous analysis will first help us to define the frequency and temperature domain where the asymptotic expression $\alpha_{Hasy.}$ is valid. If $\omega' \gg \omega_{ac}$, then the phonon population factor in Eq. 2 writes

$$n(\omega') - n(\omega' + \omega_{ac}) = \beta\hbar\omega_{ac} \frac{\exp(\beta\hbar\omega')}{(\exp(\beta\hbar\omega')-1)^2}, \tag{7}$$

with $\beta = 1/k_B T$. Integration in Eq. 1 can be done using the dimensionless quantity $z = \beta h v_{ac} x$. The temperature behavior in the asymptotic limit is obtained assuming that we can integrate $z$ from 0 up to infinity. If we want to keep a lower bound for the integration quite small while satisfying the condition $x > 20$, we should fulfill a first condition: $v_{ac}(GHz) < T(K)$. Moreover the dispersion on transverse phonon branches cannot be neglected in GaAs for frequencies above 1THz [36]. Thus temperature should be small enough to keep phonon populations above 1THz negligible. This leads to the second condition: $T < 10K$ for an error of 20% in the integration. These two conditions show that the temperature and frequency domains where $\alpha(v_{ac}) = 3.96 \times 10^{-8} v_{ac}^2 T^3$ holds are drastically limited.

We will now consider how Herring processes behave beyond these frequency and temperature ranges. A key point to answer this question is the fact that transverse phonons frequencies in GaAs have a cutoff at approximately 3THz [27]. In order to obtain the two phonon density of states

$$\rho_H(\omega_{ac}) = \int_0^{x_{max}} \rho_2(x) dx, \tag{8}$$

the integration has to be limited to a maximum value $x_{max} = 3/\nu_{ac}$ ($\nu_{ac}$ in THz), indicated by vertical dotted lines in figure 4 for different acoustic frequencies. The resulting integrated density of states $\rho_H(\omega_{ac})$ is displayed in the insert. First, it increases rapidly, and then reaches a maximum for a few hundreds of GHz and at the end drops above 1THz. It means that the contributions of Herring processes to the attenuation of subterahertz longitudinal acoustic waves should be above the asymptotic prediction given by Simons's formula, before reaching

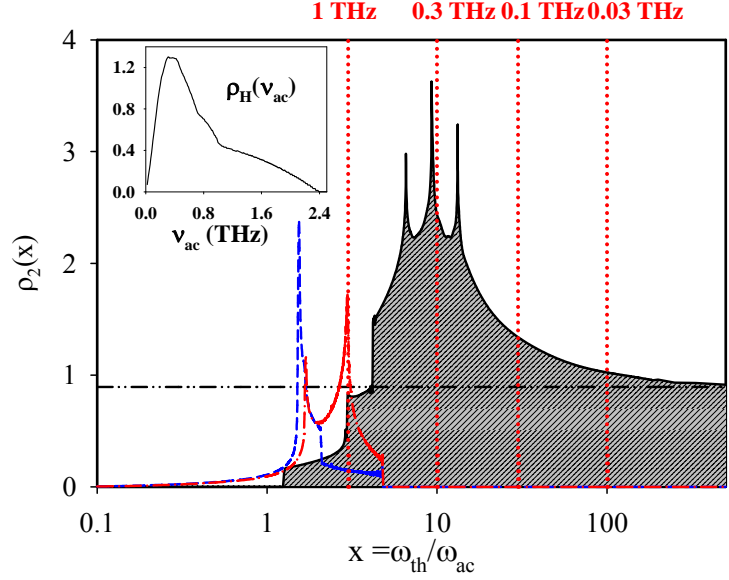

FIG. 4 (color online) Densities $\rho_2(x)$ of rescaled thermal phonons energies $x = \omega(q')/\omega_{ac}$ satisfying the scattering selection rules for the Herring processes $LA + ST \to FT$ (in full black line), neglecting phonon modes dispersion. The horizontal line gives the asymptotic value of $\rho_2(x)$. The vertical dotted lines display the integration limits to obtain the two phonon density of states for different acoustic frequencies $\nu_{ac}$ assuming a cut-off frequency of 3THz for the transverse thermal phonons; densities $\rho_2(x)$ for $LA + ST \to LA$ (dash blue line) and $LA + FT \to LA$ (dash-dot red line) processes are weaker but less affected by the cut-off frequency. The integrated two phonon density of states $\rho_H(\nu_{ac})$ for the Herring processes is displayed in the inset: the maximum is reached at 0.4 THz and the Herring processes clearly breakdown in the range 0.5 - 1THz.

a maximum. These Herring processes then breakdown above 1THz. Meanwhile the other scattering processes LA+STA→LA and LA+FTA→LA, whose contributions to $\rho_2(x)$ are displayed on fig. 4 should have a small contribution at frequencies below 0.1THz. They continuously increase with frequency and could partially compensate for the Herring processes breakdown above 1 THz. More likely, the plateau we observed experimentally is the signature of this Herring processes breakdown. Detailed calculations of scattering processes we performed in GaAs taking into account realistic phonons dispersion curves confirm this statement and moreover give the good order of magnitude for $\Delta\alpha$. This Herring processes breakdown is largely due to the characteristic shapes of transverse phonons curves in GaAs which become very flat after the half of the Brillouin zone and have a low cutoff frequency; this fact is general for crystals with long range interactions like covalent semiconductors. Thus, the same statement about Herring processes breakdown should also hold for crystals with equivalent structure and bonding like Silicon and Germanium. For example our calculations indicate that 2 phonons density of states $\rho_H$ exhibits a maximum at 0.7 THz in Si and Ge.

In conclusion, we have measured the anharmonic contribution to the mean free path of subterahertz coherent acoustic phonons in GaAs in a temperature domain much below the Debye temperature. Our experimental results show an unexpected plateau for the acoustic absorption above 0.4THz up to 1 THz. We gave qualitative arguments to demonstrate that Herring processes give the dominant contribution to this absorption up to 0.3-0.4 THz but become rapidly inefficient, due to the low frequency cutoff of transverse phonons in GaAs, giving rise to the observed plateau. This effect is responsible for the long mean free path ($l = 1/(2\alpha) = 50 \mu m$) we measured at 1 THz at 60K. An extrapolation with a $T^{-1}$ temperature dependence at room temperature leads to a MFP of 10μm which is surprisingly large but could explain the recent indirect observations of long MFP in crystalline semiconductors which have been inferred from measurements with the thermal conductivity spectroscopy technique [5]; this value is also close to the extrapolation to low-frequency modes of first principle calculations of the phonons relaxation times for normal and Umklapp scattering processes in GaAs [11] and to a rough estimation deduced from lifetime measurement of an acoustic nanocavity [37]. Thus, while the Herring processes contribution to longitudinal phonon absorption is dominant up to a few hundreds of GHz, their breakdown, which is a general statement, provides a frequency window where subterahertz coherent acoustic waves can propagate over macroscopic distances, even at room temperatures. These results are particularly appealing for phonon imaging using nanometric acoustic waves.

*Corresponding author
Bernard.perrin@insp.jussieu.fr